\documentclass[10pt, conference]{FMFP2020}
\usepackage[top=1.50cm, bottom=1.50cm, left=1.884cm, right=1.884cm, includehead, includefoot, heightrounded]{geometry}
\usepackage{fancyhdr} 
\usepackage{graphicx} 
\usepackage[textfont=bf, font = normalsize]{caption} 
\usepackage{floatrow} 
\usepackage{amssymb}
\usepackage{amsfonts}
\usepackage{amsmath}
\floatstyle{plaintop} 
\restylefloat{table} 
\usepackage{subfigure} 
\renewcommand{\headrulewidth}{0pt}

\begin{document}

\title{\LARGE{ \bf Application of machine learning for forced plume in linearly stratified medium}}
\author{\textbf{Manthan Mahajan\textsuperscript{*}, Nitin Kumar, Deep Shikha, Vamsi K Chalamalla, Sawan S Sinha}\\Department of Applied Mechanics, Indian Institute of Technology, Delhi - 110016, INDIA\\
{\textbf{*}Corresponding author}
E-mail : manthan.mahajan9596@gmail.com}

\maketitle
\thispagestyle{fancy} 
\pagestyle{plain} 

\noindent \textbf{ABSTRACT}

Direct numerical simulation (DNS) is very accurate however, the computational cost increases significantly with the increase in Reynolds number. On the other hand, we have the Reynolds Averaged Navier Stokes (RANS) method for simulating turbulent flows, which needs less computational power. Turbulence models based on linear eddy viscosity models (LEVM) in the RANS method, which use a linear stress-strain rate relationship for modelling the Reynolds stress tensor do not perform well for complex flows \cite{shih1995new} . In this work, we intend to study the performance of non linear eddy viscosity model (NLEVM) hypothesis for turbulent forced plumes in a linearly stratified environment and modify the standard RANS model coefficients obtained from machine learning. The general eddy viscosity hypothesis supported by the closure coefficients generated from the tensor basis neural network (TBNN) is used to develop TBNN based K-$\epsilon$ model. The aforementioned model is used to evaluate the plume's mean velocity profile, and maximum height reached. The comparison between standard LEVM, NLEVM and the experimental results indicates a significant improvement in the maximum height achieved, and a good improvement in the mean velocity profile.

\section{{\textbf{INTRODUCTION}}}
There are many real-life scenarios where we can see forced plumes. For example, plume formation after a volcanic eruption, fountains, pollutants flows in the atmosphere, and chimney smoke from the industries. Hence, it becomes an important aspect of engineering to simulate the flow. Also, stratified fluid studies are commonly used in geophysical fluid mechanics.

Mortan et al.\cite{morton1956turbulent} suggested in his study that plume is simply used for the flow in which there is an effect of buoyancy only, and jets are used for the flow with continuous momentum supply. Hence, "forced plume" or "buoyant jets" terms are used for the flow with both properties, simple plume and jet. Fox \cite{fox1970forced} in their study investigated the turbulent buoyant jet in a linearly stratified environment and figured out that the entrainment is dependent on the Reynolds stress, the form of similarity profiles and local mean Froude number. Bloomfield et al. \cite{bloomfield2000theoretical} in their work developed a theoretical model for an axisymmetric turbulent fountain in both homogeneous and stratified mediums. The model also incorporates the entrainment of the ambient fluid into the initial fountain up flow, and entrainment of ambient and up flow into the subsequently formed downflow. The entrainment of the denser fluid (fountain) with the ambient fluid gives two prominent effects, an increase of overall volume flux with the entry of ambient fluid and the velocity of the up-flow decreases to zero at a certain maximum height. After that up flow falls on the subsequent up flow as a turbulent plume and surrounds the central up flow. The developed model was approximately 95$\%$ accurate in calculating flow parameters in linearly stratified medium.  

Mirajkar et al. \cite{mirajkar2017effects} in their research presented the study of the effects of varying ambient stratification strength $N_\infty$ for the turbulent plume flow field. Parameters like maximum height, spreading height, and radial propagation of the plume were characterized to study the behavior of the plume. The radial spread of the plume was found to be independent of the buoyant frequency. The entrainment coefficient was found to be larger for the higher values of $N_\infty$. The maximum height of the plume and spread of the plume decreases with an increase in $N_\infty$.
Mirarjkar et al. \cite{mirajkar2020piv} studied the turbulent forced plume experimentally and compared different parameters with the jet flow field. The centre line velocity declined linearly with height until suddenly decreased to zero while the jet velocity dropped off endlessly. Although the two had different mean velocities, there was a good agreement. 

In their study, Kumar et al.\cite{kumar2022assessment} analysed and compared the RANS simulation results for mean centre line velocity and maximum height reached by the turbulent forced plume in a linearly stratified environment with the experimental results by Mirajkar et al.\cite{mirajkar2020piv}. The mean velocity profile gives accurate results in the vicinity of the source but away from the source velocity profile deviates away from the experimentally determined profile. This also leads to significant errors in the maximum height prediction by the RANS method. Various variants of the K-$\epsilon$ were analysed, and standard K-$\epsilon$ was found to be most accurate among linear eddy viscosity models.

Tsang-Hsing Shih et al. \cite{shih1995new} in their work analysed that the linear eddy viscosity models are unable to do well in flow with separation and curvatures. LEVM like standard K-$\varepsilon$ defines Reynolds stress tensor as proportional to the strain rate tensor alone. It has been observed that these models partially follow realizability conditions. The realizable algebraic equation for Reynolds stress tensor was derived in which linear to quadratic function of mean strain rate and mean rotation rate tensor were considered for defining Reynolds stress tensor. The model coefficients were derived considering all realizability conditions.

S.B. Pope \cite{pope1975more} aims to suggest a better formulation for the effective viscosity approach for Reynolds stress tensor closure. In their research, a hypothesis was proposed in which anisotropic part of Reynolds stress tensor was present as a linear summation of ten tensors. These ten tensor includes higher order terms of mean strain rate and rotation rate normalised with the k/$\varepsilon$ ratio. All tensors are multiplied with the coefficients, which are the function of a finite number of known invariants.

Bruntonv et al. \cite{brunton2020machine}  presented an article on the history, recent developments, and new opportunities in the field of machine learning when combined with the fluid mechanics problems. Gholami et al. \cite{gholami2015simulation} had compared computational fluid dynamics (CFD) and artificial neural network (ANN) methods against experimental observations studying over open channel sharp bend flow characteristics. Results indicate better performance of ANN with less root mean square (RMS) error as compared to CFD. Julia Ling et al. \cite{ling2016reynolds} justified the significant need for modification in the conventional RANS method to find Reynolds stress anisotropic tensor to define the richer set of turbulence physics. In their work, they aim to solve the Reynolds stress tensor with the help of an artificial neural network called Tensor basis neural network (TBNN). The equation used for the neural network was the same as given by Pope \cite{pope1975more}. The neural network was trained with DNS data for a simple channel flow. Reynolds stress tensor prediction by TBNN was compared with RANS (LEVM and NLEVM) and experimental data. The results were more accurate than RANS, but they could not match the level of DNS.

The previous RANS studies on the turbulent forced plume in the stratified environment were done, using simple RANS with a LEVM. No work has been recorded regarding modifying the simple RANS method and use of NLEVM for the forced plume. In the present work, we modified the RANS model with the help of a machine learning algorithm. We used the LES data set for the neural network training and applied the solution of TBNN to the realizable algebraic Reynolds stress equation model \cite{shih1995new} to compile a new TBNN based K-$\epsilon$ model. In this sense, we present a new cost-efficient RANS method incorporating a machine learning based turbulence model of the turbulent forced plume. Aforementioned tasks are achieved by working on the following objectives: 
\noindent•	Use of ML tools and high fidelity LES database to arrive at improved closure coefficients for general eddy viscosity closure paradigms which will help in developing TBNN based K-$\varepsilon$ model.
Evaluation of ML enhanced RANS methodology in simulating and predicting flow statistics in plume flow. Specifically, we look at the following quantities:
\begin{enumerate}
\item Mean velocity profile.
 \item The maximum height reached by the plume.
\end{enumerate}
\noindent All evaluations are performed by comparing our ML enhanced RANS predictions against available experimental results by Mirajkar et al. \cite{mirajkar2020piv} and RANS results by Kumar et al. \cite{kumar2022assessment}.

\section{\textbf{METHODOLOGY}}\label{sec2}
This section presents the governing equations and  problem formulation using RANS and ML methodology used for simulation of forced plume in stratified environment. Section \ref{sec2p1} presents the governing equation based on RANS and concise description of realizable algebraic Reynolds stress model \cite{shih1995new} and the standard K-$\epsilon$ model. Section \ref{2p2} presents the theory for used ML methodology and section \ref{2p3} presents the computational set up for simulations.  
\subsection{\textbf{Governing equations}}\label{sec2p1}
The unsteady Reynolds averaged Navier Stokes equations (URANS) and energy equations under the Boussinesq approximation are solved numerically using an open-source code OpenFOAM 5.0. These set of equations are given as:
\begin{equation}
    \frac{\partial\overline{u_{i}}}{\partial x_{i}} = 0
\end{equation}
\begin{equation}
    \frac{\partial\overline{u_{i}}}{\partial t} +
    \overline{u_{j}}\frac{\partial\overline{u_{i}}}{\partial x_{j}}
    = - \frac{\partial\overline{p_{d}}}{\partial x_{i}} \\ +
    \frac{\partial}{\partial x_{j}}\lbrack \nu\frac{\partial\overline{u_{i}}}{\partial x_{j}}
    - \overline{u_{i}^{'}u_{j}^{'}}{]} 
    - g_{k}x_{k}
    \frac{\partial\overline{\rho}}{\partial x_{i}}
\end{equation}
\begin{equation}
     \frac{\partial T_{d}}{\partial t} +
     \frac{\partial\overline{u_{j}}\overline{T_{d}}}{\partial x_{j}} =
     \frac{\partial}{\partial x_{j}} {[} \frac{v}{\Pr}
     \frac{\partial T_{d}}{\partial x_{j}}  -
        \overline{T'u_{j}^{'}}\rbrack{} \\- \overline{w}
     \frac{\partial T_{a}(z)}{\partial z} 
\end{equation}
\begin{equation} \label{density}
    \overline{\rho} = 1 - \beta(\overline{T} - T_b)
\end{equation}
\noindent where $\overline{\rho u_{i}^{'}u_{j}^{'}}$  represents the Reynolds stress tensor which is solved with the help of turbulence model. $\overline{p_{d}}$ is defined as $\overline{p} = \overline{p_d} + \overline{\rho}g_3x_3$. $\overline{\rho}$ is Reynolds averaged total density normalised with reference density taken at the bottom ($\rho_b$). The relation between temperature and density is give in Eq. \ref{density}. $T_{d}$ is the deviation temperature can be represented as the difference between Reynolds averaged temperature and the background temperature. Evolution equation of $T_{d}$ is found by using the equation (\ref{T_{d}}) in the evolution equation of $\overline{T}$
\begin{equation} 
\label{T_{d}}
  T_{d} = \overline{T} - T_a(z)
\end{equation}
where $\overline{T}$ is Reynolds averaged temperature and $T_a(z)$ is the background temperature varying linearly with height. A similar set of equations and the numerical setup were used by Kumar et al. \cite{kumar2022assessment} in their work to evaluate different variants of K-$\epsilon$ model for the forced plume flow in a stratified environment.
The non-linear eddy viscosity based realizable algebraic Reynolds stress model (Shih's quadratic model \cite{shih1995new}) is considered to extend the study over the turbulent forced plume. In this model, Reynolds stress depends on a combination of mean strain rate (S) and rotation rate (W) tensors up to the second order. Comparing the results for this model with the standard K-$\epsilon$ \cite{kumar2022assessment} and experimental \cite{mirajkar2020piv} is essential because the TBNN based K-$\epsilon$ model is compiled in OpenFOAM 5.0 after editing this model. The realizable algebraic Reynolds stress model directly takes the modelling equation for k (turbulent kinetic energy) and $\epsilon$ (dissipation rate) from the standard K-$\epsilon$ model, which are 
\begin{equation} 
\label{keq}
\centering
    k_{,t} + [U_jk - (\nu + \frac{\nu_t}{\sigma_k})k_{,j}]_{,j} = P + B - \varepsilon 
\end{equation}
\begin{equation}
    \varepsilon_{,t} + [U_j\varepsilon - (\nu + \frac{\nu_t}{\sigma_\varepsilon})]_{,j} = C_1\frac{\varepsilon}{k}P - C_2\frac{\varepsilon^2}{k} \\ + C_1(1-C_3)\frac{\varepsilon}{k}B
\end{equation}
\begin{equation} 
\label{cmu}
\text{where}\hspace{0.5cm} \nu_t = C_{mu}\frac{K^2}{\varepsilon}, C_{mu} = \frac{2/3}{A_1+\eta+\alpha\zeta} ,\\
\zeta = \frac{K}{\varepsilon}W
\end{equation}
$\nu_t$ is turbulent viscosity. {$( )_{,t}$} represents derivative of ( ) with respect to time and {$( )_{,i}$} represents spacial derivative of ( ) with respect to $x_i$. P is the turbulence production term due to shear, while B is the turbulence production term due to buoyancy.  Standard and machine learning based models are modified to account for this aspect.

B is modeled using gradient diffusion hypothesis as in \cite{kumar2022assessment}, according to which it is dependent on the total Reynolds stress.
\begin{equation} 
\label{buoyancy production}
    B = (\beta)(\frac{2}{3})(\frac{C_{mu}}{\sigma_t})(\frac{k}{\varepsilon})(\overline{u_iu_j})\frac{\partial\overline{T}}{\partial x_j}g_j
\end{equation}

Other than the order of mean strain rate and rotation rate tensor, another significant difference between algebraic stress and the standard K-$\epsilon$ model is the factor $C_{mu}$. This factor is constant in the standard K-$\epsilon$ model and variable in the algebraic stress model to follow the realizability criteria defined by Shih et al. \cite{shih1995new}.

\subsection{\textbf{Machine learning methodology}}\label{2p2}
Integration of the standard RANS method with machine learning has been done with the help of a tensor basis neural network (TBNN). TBNN was formulated by Ling et al. \cite{ling2016reynolds}, which is an artificial neural network. The purpose of this network is to predict the Reynolds stress anisotropic tensor by taking highly accurate data for training. The Galilean invariance is embedded into the predicted Reynolds stress anisotropic tensor. The purpose for embedding the Galilean invariance is to ensure that the anisotropic tensor is rotated by the same amount when the coordinate frame is rotated. Without invariance property, the machine learning model will make different predictions of the same flow if its direction is changed.
\begin{figure}
\centering
\includegraphics[width=\textwidth]{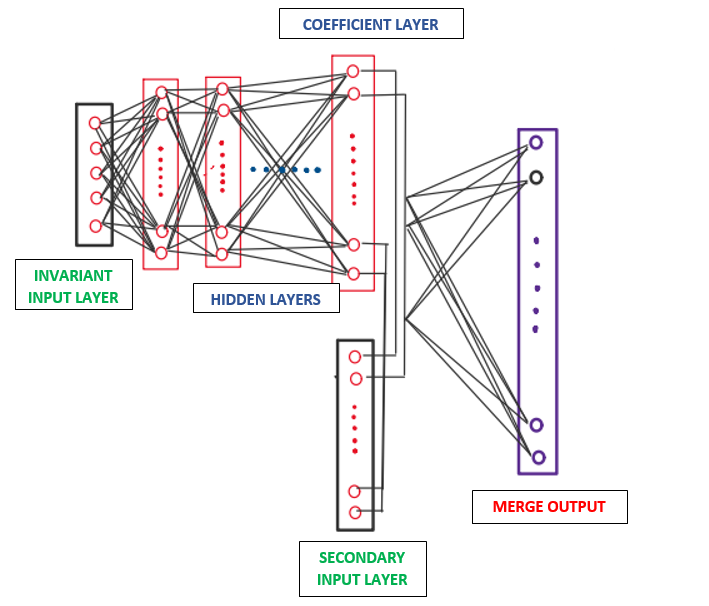}
\caption{TBNN Architecture.\cite{ling}}
\label{figure1}
\end{figure} 
\noindent Fig. \ref{figure1} shows the basic structure of TBNN. There are two input layers. One is for the five tensor invariants, and the other is used to input the ten tensor basis. The first input layer is followed by some optimised number of hidden layers, which apply the activation function to the input parameters. The final hidden layer has ten nodes representing ten values of $g^{(n)}$ which will be merged with ten tensor input layer $T^{(n)}$ to give the output as 
\begin{equation} \label{anisotropic}
    b=\sum_{n = 1}^{10}g^{(n)}(\lambda_{1},\lambda_{2},\ldots\ldots, \lambda_{5})
T^{(n)}
\end{equation}
where b is the normalised anisotropic Reynolds stress tensor. The equation \ref{anisotropic} for Reynolds stress anisotropic tensor was given by Pope \cite{pope1975more}. Five tensor invariants
$\lambda_{1},\lambda_{2},\ldots\ldots, \lambda_{5}$ are the scalar functions of the normalised mean strain rate tensor ($S_n$) and rotation rate tensor ($W_n$). The tensors of $T^{(n)}$ (n = 1 to 10) are represented in terms of $S_{n}$ and $W_{n}$ as\\
\begin{align}
\label{tensors}
T^{(1)} = S_{n} \nonumber\\ T^{\left( 2 \right)} = S_{n}W_{n} - W_{n}S_{n} \nonumber\\ T^{\left( 3 \right)} = S_{n}^{2} - 1/3 I. Tr (S_{n}^{2}) \nonumber\\ T^{\left( 4 \right)} = W_{n}^{2} - 1/3 I. Tr (W_{n}^{2}) \nonumber\\ T^{\left( 5 \right)} = W_{n}S_{n}^{2} - S_{n}^{2}W_{n} \nonumber\\
T^{\left( 6 \right)} = W_{n}^{2}S_{n} + S_{n}W_{n}^{2} -2/3I. Tr(S_{n}W_{n}^{2} \nonumber\\ T^{\left( 7 \right)} = W_{n}S_{n}W_{n}^{2} - W_{n}^{2}S_{n}W_{n} \nonumber\\ T^{\left( 8 \right)} = S_{n}W_{n}S_{n}^{2} - S_{n}^{2}W_{n}S_{n} \nonumber \\ T^{\left( 9 \right)} = W_{n}^{2}S_{n}^{2} + S_{n}^{2}W_{n}^{2} -2/3I. Tr(S_{n}^{2}W_{n}^{2}) \nonumber\\ T^{(10)} = W_{n}S_{n}^{2}W_{n}^{2} - W_{n}^{2}S_{n}^{2}W_{n} \nonumber \\
\end{align}
Various scalar invariants functions can be given as \\
\begin{align}
\lambda_{1} = Tr (S_{n}^{2}) \nonumber\\    
\lambda_{2} = Tr(W_{n}^{2}) \nonumber\\    
\lambda_{3} = Tr(S_{n}^{3}) \nonumber\\
\lambda_{4} =Tr(W_{n}^{2}S_{n}) \nonumber\\    
\lambda_{5} = Tr(W_{n}^{2}S_{n}^{2}) \nonumber\\
\end{align}
Calculating the values of the scalar coefficients $g^{(n)}(\lambda_{1},\lambda_{2}, .~.~.~, \lambda_{5})$ is the main purpose of the neural network. These coefficients are used in the non linear stress equation to make a new turbulence model in OpenFoam $5.0$.
\subsection{\textbf{Computational methodology}} \label{2p3}
\begin{figure}
\centering
\includegraphics[width=0.8\textwidth]{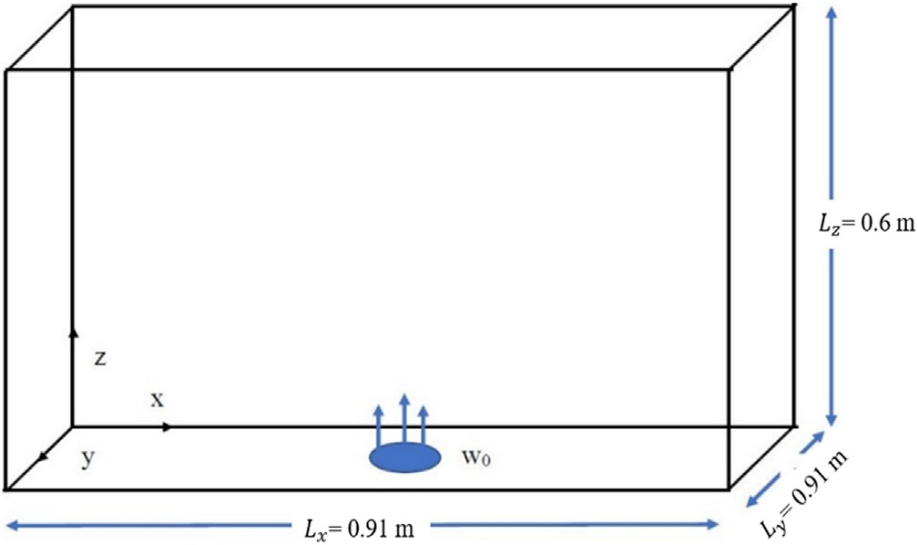}
\caption{Computational domain.}
\label{figure2}
\end{figure}

The computational domain under the study is a three dimensional box with dimensions as  $L_x$ = 0.91m, $L_y$ = 0.91m and $L_z$ = 0.6m with a source of plume at the center indicated by a patch as shown in Fig. 
\ref{figure2} 
To compare the results with the experimental data of Mirajkar et al. \cite{mirajkar2020piv} and RANS based data of standard K-$\epsilon$ for the turbulent forced plume, the same flow parameters are used in the study. $w_{o}$ = 0.22 $ms^{-1}$ is the initial velocity in the vertical direction from the source, which corresponds to Re = 3100. All other components of velocity are zero. The stratification strength of the ambient fluid ($N_{\infty}=0.4 s^{-1}$) is used. A constant temperature $T = 316 K$ is used at the source. The diameter ($d$) of the source is 0.0127m. At the source, a fixed velocity is used as the boundary condition, and the no-slip boundary is applied at the bottom boundary. The domain's lateral and top boundaries are treated with a zero-gradient boundary condition. At the source, a constant temperature is used, while the zero gradient is used at all other boundaries.

LES data for turbulent plume is used for the training of the TBNN. The overall domain consists of approximately 7 million grid points. This large number of data points can slow the training of the neural network and may also lead to overfitting and generation of bias, as most of the data have negligible information about the flow characteristics. We choose a subdomain to pick the data points that can significantly describe the turbulent plume's behaviour. Finally, the data set is reduced to 1.7 million.

We tested different configurations of the neural network and its parameters for the sample data set, which is a proportional extraction of the complete data set. The change in learning rate does not affect the loss versus epoch curves. The larger batch sizes are used to change the higher learning rate curve to a good learning rate curve. It is also observed that increasing the number of hidden layers and the number of neurons per hidden layer helps decrease the RMS error value between the predicted and true value of the Reynolds stress anisotropic tensor. The comparison for the selection of the neural network configuration is made based on the root mean square (RMS) error value as in table \ref{table1} (numerically) and with the prediction plot over the test data (visually) in Fig. \ref{figure4}.

\begin{figure}
\centering
\includegraphics[width=0.8\textwidth]{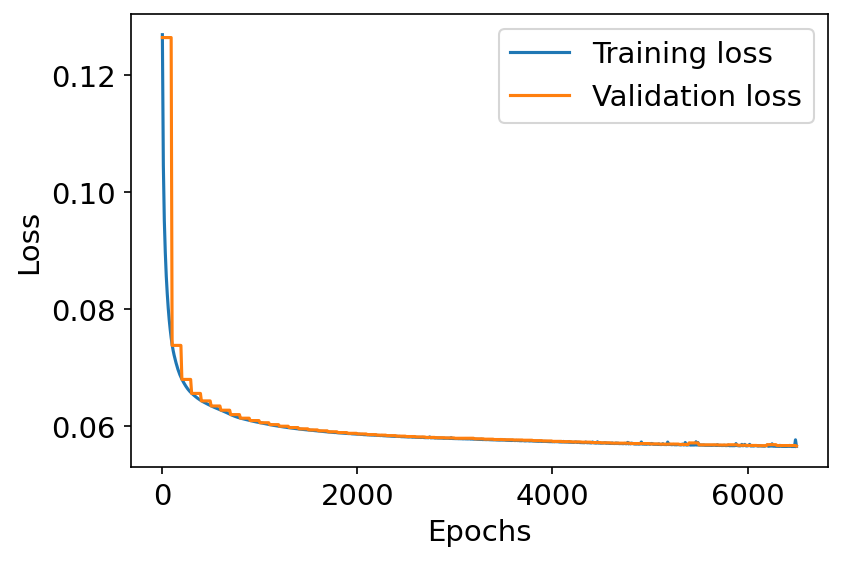}
\caption{Loss versus epochs curve.}
\label{figure3}
\end{figure}

\begin{figure}
\centering
\includegraphics[width=0.9\textwidth]{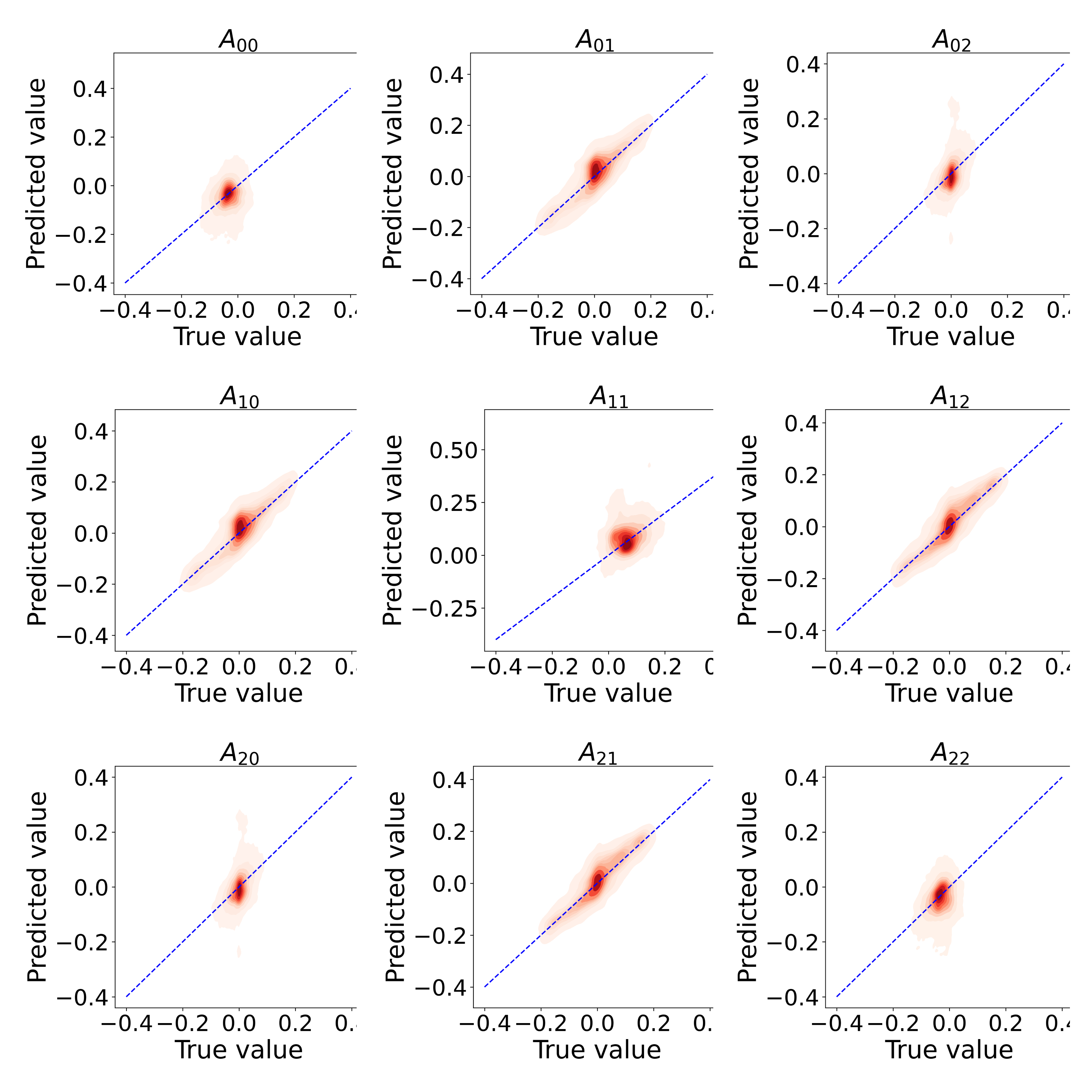}
\caption{Output prediction plot over test data}
\label{figure4}
\end{figure}

Network 1 shown in table \ref{table1} is 24 hidden layers with 88 nodes per hidden layer with 55000 batch size; Network 2 is 24 hidden layers with 104 nodes with 65000 batch size, and Network 3 is 32 hidden layers with 128 nodes in each hidden layer with 80000 batch size. Network 2 is selected as it has less RMS error for the diagonal components and approximately similar error for the diagonal elements as compared to Network 1. 
\begin{table}
\centering
\caption{The RMS error value for different networks}
\begin{tabular}{|c|c|c|c|c|}
\hline
 Stress  & network 1 & network 2 & network 3 \\ 
 components &  & &\\
\hline
 $A_{00}$&$0.0592$  &$0.0583$ &$0.0606$\\
$A_{01}$&$0.0514$  &$0.0518$ &$0.0520$\\
$A_{02}$&$0.0571$  &$0.0575$ &$0.0579$\\
$A_{10}$&$0.0514$  &$0.0518$ &$0.0520$\\
$A_{11}$&$0.0765$  & $0.0752$ &$0.0797$\\
$A_{12}$&$0.0474$  &$0.0478$ &$0.0483$\\
$A_{20}$&$0.0571$  &$0.0575$ &$0.0579$\\
$A_{21}$&$0.0474$  &$0.0478$ &$0.0483$\\
$A_{22}$&$0.0568$  &$0.0559$ &$0.0584$\\
\hline
\end{tabular}
\label{table1}
\end{table}
For the network with 24 hidden layers and 104 nodes per hidden layer, the loss versus epoch curve and output prediction plot for test data are shown in Fig. \ref{figure3} and Fig. \ref{figure4}, respectively. Fig. \ref{figure3} shows the convergence of training and validation loss, representing a good fit curve. Fig. \ref{figure4} shows the alignment of true and predicted data to a diagonal line with a slope as one. Over this line, the true and predicted values are equal. The important hyper parameters which are used to optimise the TBNN algorithm for forced plume data set are in table \ref{table2}.

\begin{table}
\centering
\caption{Hyper parameters used  in TBNN }
\begin{tabular}{|c|c|}
\hline
Number of hidden layers & 24 \\ 
\hline
Number of nodes per & 104\\ hidden layer &  \\
\hline
Loss function & Mean square error  \\
\hline
Total data set & $17.5 \times 10^5$\\
\hline
Split fraction & 0.7  \\
\hline
Optimiser & ADAM \\
\hline
Batch size & 65000  \\
\hline
Learning rate & 0.01  \\
\hline
\end{tabular}
\label{table2}
\end{table}

\section{\textbf{RESULTS AND DISCUSSION}}\label{sec3}
A grid independent mesh of 1.34 million cells used by Kumar et al. \cite{kumar2022assessment} previously in their work for analysing different variants of K-$\epsilon$ model has been utilized. The stratification strength $N_\infty$ used is $0.4 s^{-1}$. The data for same value of $N_\infty$ is used for the training of TBNN algorithm. We modified the standard Shih's quadratic model to develop a new turbulence model for the turbulent forced plume referred to as TBNN based K-$\varepsilon$ model. The non-linear stress equation is coded in reference to the Pope \cite{pope1975more} in this model. The ML coefficients, generated from the TBNN code after training the neural network with the generated LES data, are used as the closure coefficients for solving the Reynolds stress equation. The simulations are performed in Open FOAM 5.0. The modified BuoyantBousssinesqPimpleFoam is used for the simulations. The Solver is modified by Kumar et al. \cite{kumar2022assessment} to capture ambient stratification's effects.

Simulations are performed for a time of 160 seconds. Results are time-averaged from the span of 140 seconds to 160 seconds. The comparisons of the results  were made obtained from the Shih's quadratic model, the standard K-$\varepsilon$ model by Kumar et al. \cite{kumar2022assessment}, the experimental results by \cite{mirajkar2020piv} and the TBNN based K-$\varepsilon$. In the comparison, we have considered the mean velocity profile and the maximum height reached by the plume. The maximum height of the plume is the height where the mean velocity becomes zero. All plots drawn represent the normalised mean velocity versus the normalised height. The vertical centre line mean velocity ($<U_y>$) is normalised using the initial source velocity ($w_o$), and the height (Z) is normalised using the source diameter (d).
\begin{figure}
\centering
\includegraphics[width=0.9\textwidth]{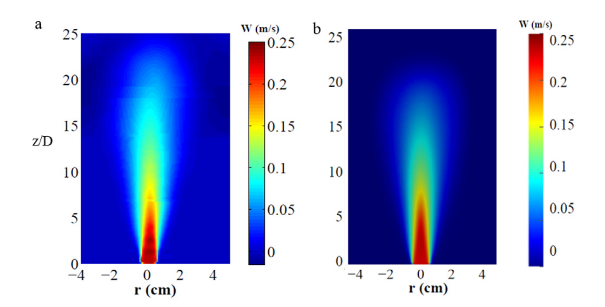}
\caption{Mean vertical velocity contour with normalised height(Z/d)
 (a) Experimental\cite{Mirajkar} (b) K-$\varepsilon$ model\cite{KUMAR}}
\label{figure5}
\end{figure}
\begin{figure}
\centering
\includegraphics[width=0.9\textwidth]{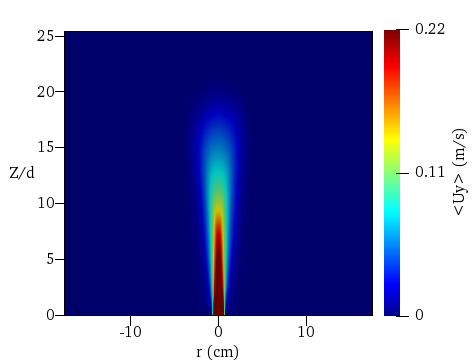}
\caption{Shih's quadratic model mean vertical velocity contour plot}
\label{figure6}
\end{figure} 
When released from the source at a higher temperature and lesser density than the ambient environment, the plume rises due to both momentum and buoyancy effects. The ambient environment has varying density. Near the source, momentum effects are dominant, while away from the source, buoyancy effects are dominant. As the plume begins to rise, there is constant entrainment of the cold fluid into the plume from the ambient. This process makes the plume denser while rising. The height where the plume's density becomes equal to the ambient density is known as neutral height. The maximum height reached by the plume is greater than the neutral height. Due to the inertia in plume surpasses the neutral height and then falls back.
\begin{figure}
\centering
\includegraphics[width=0.9\textwidth]{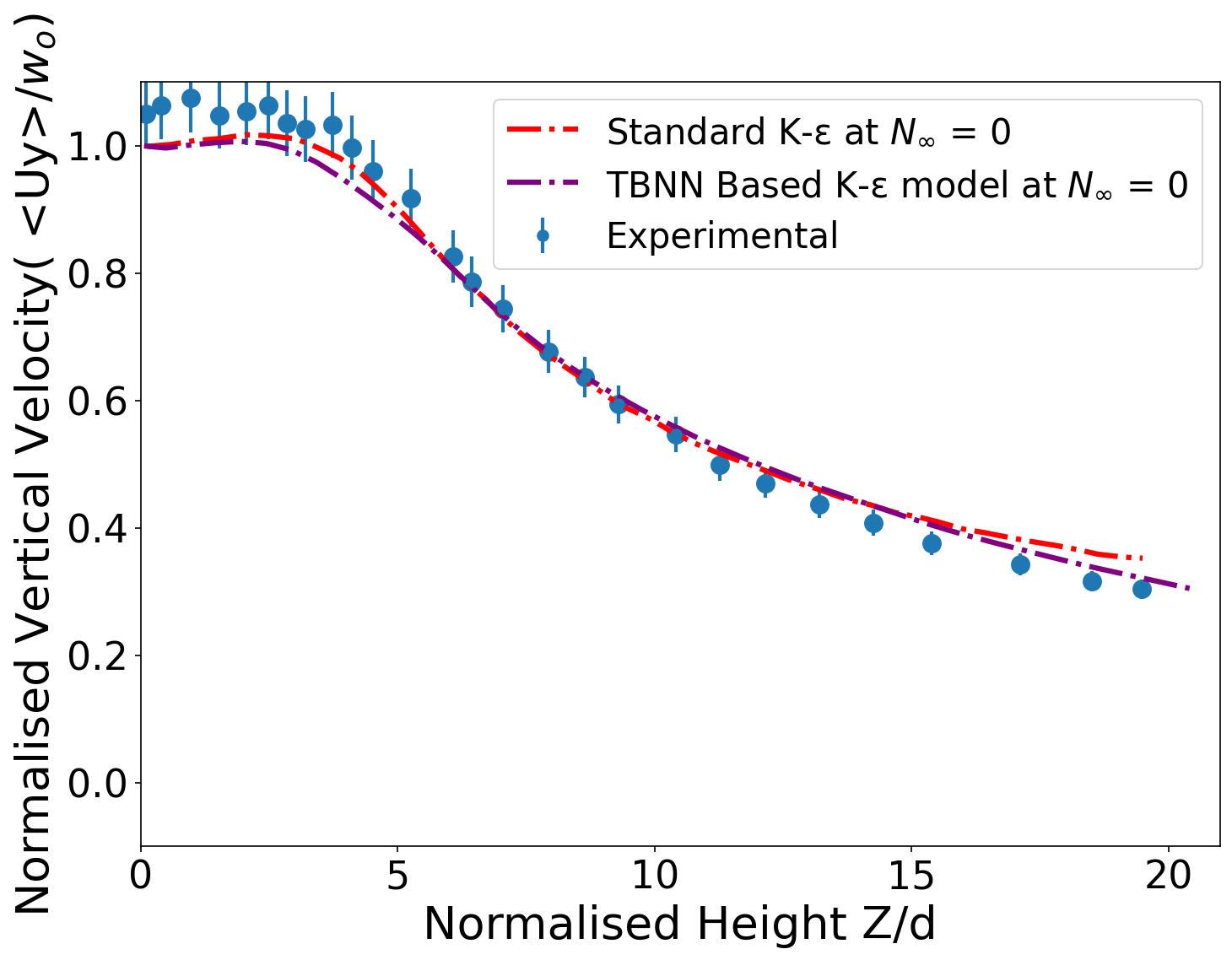}
\caption{Normalised mean center line velocity profile comparison between standard K-$\varepsilon$, TBNN based K-$\varepsilon$ model and Experimental results for $N_\infty$ = 0}
\label{figure7}
\end{figure}

Fig. \ref{figure5} and \ref{figure6} represent the mean vertical velocity contour plots of the turbulent forced plume obtained from the experimental data of Mirajkar et al \cite{mirajkar2020piv}, standard K-$\epsilon$ model, and Shih's quadratic model. The mean vertical velocity from the two models is compared with that of experimental data. It can be observed that the red color portion in mean vertical velocity contour plot of Shih's quadratic model is present up to higher height than that of the experimental and standard K-$\varepsilon$ model. Hence, the mean velocity near the source is over predicted for the quadratic model. 
\subsection{\textbf{Validation of the model}}
The TBNN algorithm is trained with LES data for the turbulent forced plume with a stratification strength ($N_\infty$) of $0.4 s^{-1}$. For this $N_\infty$ value, there was an overall improvement in predicting the results for mean velocity profile and maximum height reached by plume with the TBNN based K-$\varepsilon$ model. To validate our model, we used the same TBNN based K-$\varepsilon$ model, which we formulated for flow with $N_\infty = 0.4 s^{-1}$ to the turbulent forced plume with zero stratification strength $N_\infty = 0$ . In the Fig. \ref{figure7}, we can observe that TBNN based K-$\varepsilon$ model seems to validate the experimental results for the case of $N_\infty = 0$ flow field. The standard K-$\varepsilon$ model also predicts the mean velocity profile with reasonable accuracy. Away from the source TBNN based K-$\varepsilon$ agrees with experimental results more than standard K-$\varepsilon$, which seems to diverge from the experimental data.
\vspace{0.2cm}
 \begin{figure}
\centering
\includegraphics[width=0.9\textwidth]{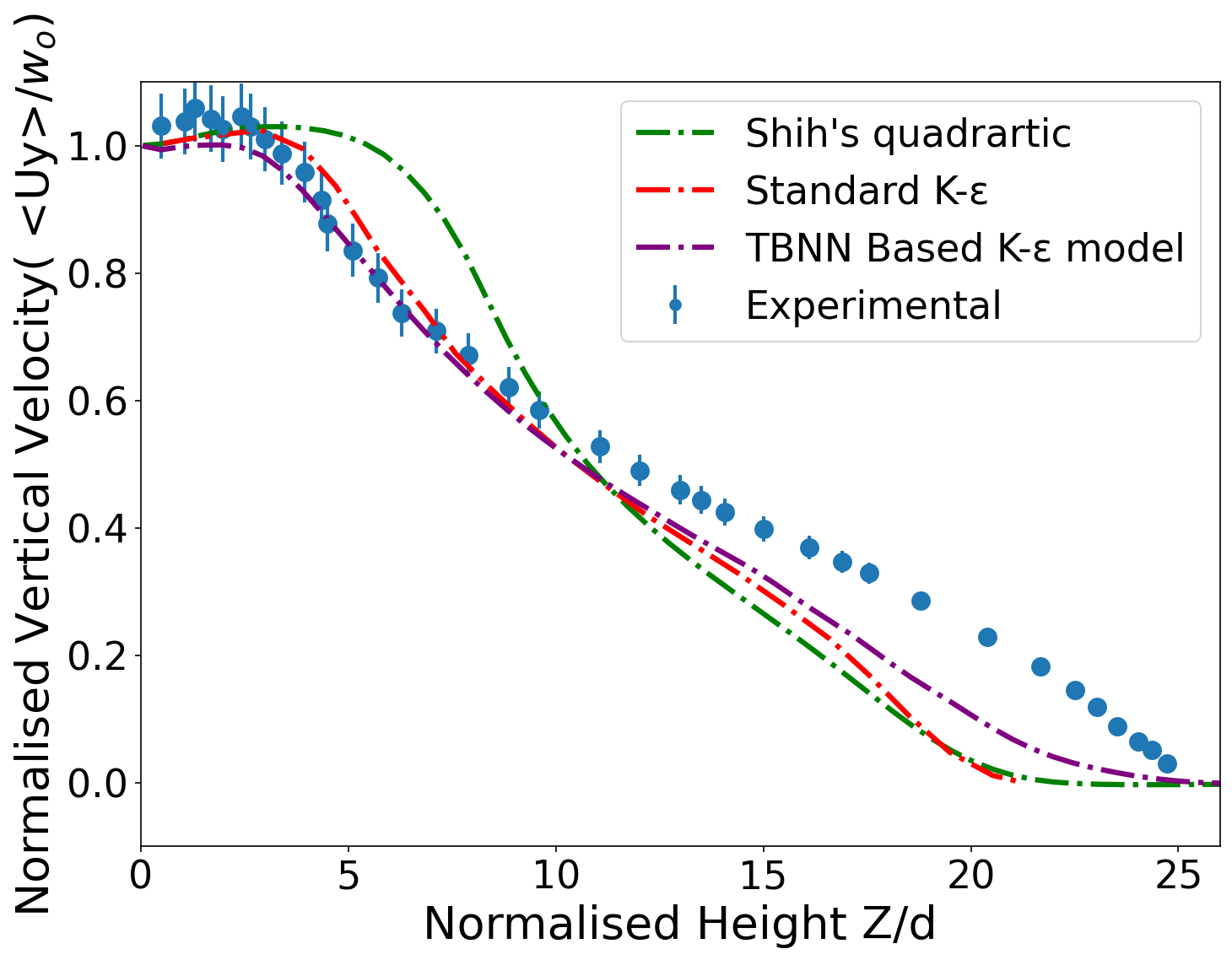}
\caption{Normalised mean center line velocity ($<U_y>/w_o$) versus normalised height(Z/d) comparison for standard K-$\varepsilon$, Shih's quadratic model, TBNN based K-$\varepsilon$ and Experimental for $N_{\infty}$ = 0.4 $s^{-1}$. The error bar represents that values can vary in this range.}
\label{figure8}
\end{figure} 
\noindent The observations made from the contour plots in Fig. \ref{figure5} and \ref{figure6} can also be verified from the normalised centre line velocity plot shown in Fig. \ref{figure8}. The standard Shih's quadratic model predicts higher values for the mean velocity in the vicinity of the source. The centre line mean velocity contour plot for TBNN based K-$\epsilon$ model is shown in Fig. \ref{figure9}. This contour plot matches the experimental data contour more than any standard turbulence model. The TBNN based K-$\epsilon$ model has significantly improved overall results for the mean velocity profile and maximum height reached by the plume. The maximum height predicted by the linear and quadratic models shown in Table \ref{table3} has an error up to 14 percent when compared with the experimental maximum height. However, the maximum height prediction by TBNN based K-$\epsilon$ is very close to the experimental results with an error of 0.26 percent shown in Table \ref{table3}.
\begin{figure}
\centering
\includegraphics[width=0.9\textwidth]{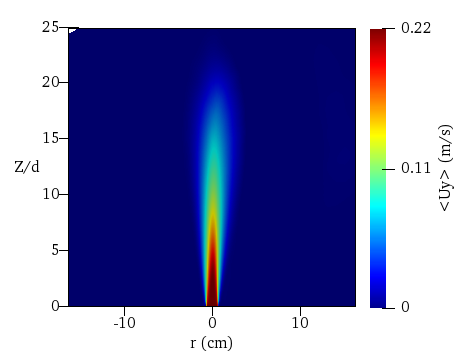}
\caption{Mean vertical velocity contour plot for TBNN based K-$\varepsilon$}
\label{figure9}
\end{figure}
\begin{table}
\centering
\begin{tabular}{ |c|c|c| }
\hline
Turbulence Models  & Maximum Height  & $\%$ error over \\
~ & Attained & experimental \\
\hline
  Standard K-$\varepsilon$ & 26.87 & 13.99 $\%$ \\ 
  \hline
  Shih's quadratic model &  27.36 & 12.42 $\%$\\ 
  \hline
  Experimental \cite{Mirajkar}& 31.24 & - \\
  \hline
  Theoretical Height \cite{morton1956turbulent} & 30.60 & 2.05 $\%$\\
  \hline
  TBNN based K-$\varepsilon$ & 31.32 & 0.26 $\%$\\
  \hline
\end{tabular}
  \caption{Maximum height (in cm) attained by plume at 0.4 stratification strengths.}
 \label{table3}
\end{table}
\section{\textbf{CONCLUSIONS}}\label{sec4}
In this paper, we present the modified RANS model for the turbulent forced plume using the TBNN based turbulence model. The evaluation of different neural network configurations was done based on the RMS error between true and the predicted value of the Reynolds stress tensor for the test data. The LES data of forced plume stratification strength of 0.4 $s^{-1}$ was used for training and testing the neural network. The coefficients generated from the TBNN were implemented in the K-$\epsilon$ model to compile a new TBNN based K-$\epsilon$ model. The evaluation of TBNN based K-$\epsilon$ model significantly improved the normalised mean velocity profile and maximum height prediction of the forced plume at $N_\infty = 0.4 s^{-1}$ . The TBNN based K-$\epsilon$ was tested on the flow with $N_\infty = 0$ and the agreement was very good with the experimental data as well as the standard K-$\epsilon$ model results.
\bibliographystyle{amsplain}
\bibliography{FMFP2020}    
\end{document}